# Anharmonicity strongly enhancing thermal interface conductance: A new anharmonic atomistic Green's function formalism

Jinghang Dai and Zhiting Tian[*]

*Sibley School of Mechanical and Aerospace Engineering, Cornell University, NY 14853, USA*
[*] *Address all correspondence to Zhiting Tian's E-mail:* [zhiting@cornell.edu](mailto:zhiting@cornell.edu)

**Abstract:** The traditional atomistic Green's function (AGF) was formulated in the harmonic regime, preventing it from capturing the role of anharmonicity in interfacial thermal transport. Incorporating anharmonicity into AGF has long been desired but remains challenging. We developed a rigorous anharmonic AGF model to incorporate anharmonicity at interfaces in 3-D structures with first-principles force constants. Thermal conductance of silicon- and aluminum-based interfaces is significantly enhanced resulting from the new channels opened by inelastic scattering. This work represents a major step forward for AGF and highlights the importance of anharmonicity at interface.

*Introduction*—Heat dissipation in microelectronics is a burning issue that limits their performance and reliability[1]. Large resistance at solid-solid interfaces often presents the major bottleneck for heat removal[2]. Understanding interfacial thermal transport and engineering interfaces with ultrahigh interfacial thermal conductance are, thus, in great demand. Atomistic Green's function (AGF) has been a powerful tool to study nanoscale thermal transport[3,4], especially across interfaces. The traditional AGF[3,5–12] was formulated within the harmonic regime and the lack of anharmonicity has become a major limitation for AGF. Including anharmonicity in AGF is in principle possible, but very challenging. Since Mingo included anharmonicity for a 1-D atomic junction in 2006[13], there have been few attempts to include anharmonicity for 3-D structures using different levels of approximation, such as obtaining the phonon inelastic scattering through a semi-empirical way[14–17]. A general and rigorous anharmonic AGF formalism is lacking.

In this Letter, we developed a general anharmonic AGF formalism to rigorously treat the phonon-phonon scattering processes at the interface using first-principles force constants for 3-D structures. To include many-body interactions via Keldysh formalism[18–21], we formed a new Fourier decomposition method for 3$^{rd}$ order tensors and defined the many-body self-energy for 3-D structure in reciprocal space. By employing this method on silicon/germanium (Si-/Ge) interfaces with first-principles force constants, we demonstrated that the presence of anharmonicity at the interface can significantly enhance interfacial thermal transport. Moreover, we showed that the substantial enhancement of the interfacial thermal conductance holds over a wide range of mass ratios, providing universal evidence of the important role anharmonicity plays at the interface. The 3-D anharmonic AGF method we present here paves the way for AGF studies of interfacial nanoscale thermal transport when anharmonicity is not negligible.

*Method derivation*—We begin by introducing the basics of traditional AGF. Typically, the system is composed of two semi-infinite leads and one central region. Retarded Green's function, $\boldsymbol{G}^r$, describes the dynamics of phonons in the center region, taking the effect of the leads into account through self-energies:

$$\boldsymbol{G}^r = [\omega^2 \boldsymbol{I} - \boldsymbol{H}_C - \boldsymbol{\Sigma}_L^r - \boldsymbol{\Sigma}_R^r]^{-1} \tag{1}$$

where $\boldsymbol{\Sigma}_{L(R)}^r \equiv \boldsymbol{H}_{CL(CR)} \boldsymbol{g}_{L(R)}^r \boldsymbol{H}_{LC(RC)}$. $\boldsymbol{H}_{CL(CR)}$, $\boldsymbol{H}_{LC(RC)}$ are harmonic force constant matrices connecting the left or right lead to the center, and $\boldsymbol{g}_{L(R)}^r$ is the uncoupled retarded Green's function for the semi-infinite leads. For detailed information, please refer to Ref [3].

To account for the contribution from three-phonon scattering, we added the many-body self-energy, $\boldsymbol{\Sigma}_M^r$, into the total self-energy[13]: $\boldsymbol{G}^r = [\omega^2 \boldsymbol{I} - \boldsymbol{H}_C - \boldsymbol{\Sigma}_L^r - \boldsymbol{\Sigma}_R^r - \boldsymbol{\Sigma}_M^r]^{-1}$. $\boldsymbol{\Sigma}_M^r$ can be obtained from the coupled non-equilibrium Green's functions $\boldsymbol{G}^<$, $\boldsymbol{G}^>$, or lesser, greater Green's function, and the 3$^{rd}$ order anharmonic tensor $\boldsymbol{\mathcal{V}}_{ijk}$ ($i$, $j$, $k$ are the indices for 3$^{rd}$ order tensor). For non-equilibrium Green's function calculation, the temperature at both leads are different and the temperature information of the leads represented by Bose-Einstein distribution for phonons are already included in the uncoupled Green's functions: $\boldsymbol{g}_{L(R)}^<$ and $\boldsymbol{g}_{L(R)}^>$.

A major challenge in implementing anharmonic AGF for a 3-D system is that the sizes of all matrices and tensors are so large that one cannot directly deal with. Fourier decomposition was applied to investigate the harmonic phonon transmission in carbon nanotube junctions[10]. In 3-D structures, the harmonic matrices can be written as $H(\vec{R}_a, \vec{R}_b)$, where $\vec{R}_a$ and $\vec{R}_b$ are the position vectors for the unit cells. By defining a unitary $\boldsymbol{P}$ matrix[10] as $P(\vec{R}_a, \vec{Q}_\alpha) = \frac{1}{N} I_H e^{i\vec{R}_a \cdot \vec{Q}_\alpha}$, we extend the Fourier decomposition to 3-D structures via the equation $\widetilde{\boldsymbol{H}} = \boldsymbol{P}^{-1} \boldsymbol{H} \boldsymbol{P}$, where N is the number of unit cell along transverse direction. The only remaining submatrices in $\widetilde{\boldsymbol{H}}$ will be the diagonal ones $\widetilde{H}(\vec{Q}_\alpha)$:

$$\widetilde{H}(\vec{Q}_\alpha) = \sum_{a=1}^{N^2} \sum_{b=1}^{N^2} H(\vec{R}_a, \vec{R}_b) e^{-i\vec{Q}_\alpha \cdot (\vec{R}_a - \vec{R}_b)} \qquad (2)$$

where $\vec{Q}_\alpha$ is the transverse wave vector. Through this method, all the Green's function and self-energy matrices only need one wave vector $\vec{Q}_\alpha$ to represent in reciprocal space.

The only exception, however, is the anharmonic tensor which is defined in 3-D format and cannot be "diagonalized" via the matrix formulation. Well-established tensor decomposition methods[22] such as Tucker decomposition/higher-order singular value decomposition (HOSVD) or canonical/parallel factor (CP) decomposition all fall short of reducing the dimension of the 3rd order tensor into a matrix form with the Fourier-component patterns[23]. Therefore, we developed a completely new method from scratch.

For any anharmonic tensor $\boldsymbol{\mathcal{V}}$ or $\mathcal{V}(\vec{R}_a, \vec{R}_b, \vec{R}_c)$, the 3rd order tensor Fourier decomposition in 3-D structure will be:

$$\widetilde{\mathcal{V}}_{uvw} = \sum_{ijk} \mathcal{V}_{ijk} P_{ui}^{-1} P_{vj}^{-1} P_{wk} \text{ or } \widetilde{\mathcal{V}}_{pqr} = \sum_{lmn} \mathcal{V}_{lmn} P_{pl}^{-1} P_{qm} P_{rn} \qquad (3)$$

By doing so, we only need two wave vectors, $\vec{Q}_\psi$ and $\vec{Q}_\theta$, to represent such 3rd order tensor $\mathcal{V}(\vec{R}_a, \vec{R}_b, \vec{R}_c)$ in real space as $\widetilde{\mathcal{V}}(\vec{Q}_\psi, \vec{Q}_\theta)$ in reciprocal space.

Accordingly, we defined the lesser(greater) many-body self-energy in reciprocal space as:

$$\widetilde{\Sigma}_{M,u,r}^{<(>)}(\omega, \vec{Q}) = i\hbar \int_{-\infty}^{\infty} \sum_{\vec{Q}'} \sum_{vwpq} \left[ \widetilde{\mathcal{V}}_{uvw}(\vec{Q}, \vec{Q}') \widetilde{G}_{wp}^{<(>)}(\omega - \omega', \vec{Q}') \mathcal{V}_{pqr}(\vec{Q}, \vec{Q}') \widetilde{G}_{qv}^{<(>)}(\omega', \vec{Q}') \right] d\omega'$$

(4)

This Fourier decomposition method allows us to rigorously incorporate the anharmonic force constants into AGF for 3-D structures. After a careful derivation of every single equation, concise analytic expressions for computing many-body Green's function and self-energy matrices in reciprocal space are given as:

$$\widetilde{G}^{<(>)}(\omega, \vec{Q}) = \widetilde{G}^r(\omega, \vec{Q}) \widetilde{\Sigma}^{<(>)}(\omega, \vec{Q}) \left(\widetilde{G}^r(\omega, \vec{Q})\right)^+ \qquad (5)$$

$$\widetilde{G}^r(\omega, \vec{Q}) = \left[\omega^2 I - \widetilde{H}_C(\vec{Q}) - \widetilde{\Sigma}_L^r(\omega, \vec{Q}) - \widetilde{\Sigma}_R^r(\omega, \vec{Q}) - \widetilde{\Sigma}_M^r(\omega, \vec{Q})\right]^{-1} \qquad (6)$$

$$Im[\widetilde{\Sigma}_M^r(\omega, \vec{Q})] = \frac{\widetilde{\Sigma}_M^>(\omega, \vec{Q}) - \widetilde{\Sigma}_M^<(\omega, \vec{Q})}{2i}, Re[\widetilde{\Sigma}_M^r(\omega, \vec{Q})] = \int_{-\infty}^{\infty} p.v.\left(\frac{1}{\omega - \omega'}\right) \times \frac{-1}{\pi} Im[\widetilde{\Sigma}_M^r(\omega', \vec{Q})] d\omega'$$

$$\tag{7}$$

where *p.v.* means Cauchy principle value. Together with Eq.(4), a self-consistent calculation[13,19] is required for obtaining $\tilde{G}^{<(>)}$, from which we can calculate the heat current density on the left or right side of the interface:

$$J_{L(R)}(\omega) = +(-)\sum_{\vec{Q}} \text{Tr}[\tilde{\Sigma}^{>}_{L(R)}(\omega,\vec{Q})\tilde{G}^{<}(\omega,\vec{Q}) - \tilde{\Sigma}^{<}_{L(R)}(\omega,\vec{Q})\tilde{G}^{>}(\omega,\vec{Q})]\frac{\hbar\omega}{2\pi} \tag{8}$$

The integral of current density over frequency yields the total heat current, which is required to determine two-probe interfacial thermal conductance:

$$\sigma = \lim_{T_L \to T_R} \frac{1}{T_L - T_R} \frac{1}{N^2 A_s} \int_0^\infty d\omega \sum_{\vec{Q}} \text{Tr}[\tilde{\Sigma}^{>}_{L(R)}(\omega,\vec{Q})\tilde{G}^{<}(\omega,\vec{Q}) - \tilde{\Sigma}^{<}_{L(R)}(\omega,\vec{Q})\tilde{G}^{>}(\omega,\vec{Q})]\frac{\hbar\omega}{2\pi} \tag{9}$$

*Results and Discussions*—To demonstrate the impact of anharmonicity at the interface, we performed simulations for Si/Ge interface in the [100] direction, using the force constants obtained from first-principles density-functional theory (DFT) calculations. The 2$^{nd}$ order force constants were validated by comparing the phonon dispersion with experimental dispersion[24]. The 3$^{rd}$ order ones were validated by solving the phonon Boltzmann transport equation using ShengBTE[25] and a good agreement is observed (161 W/mK at 300K) compared with the experimental data[26] (155 W/mK at 300K).

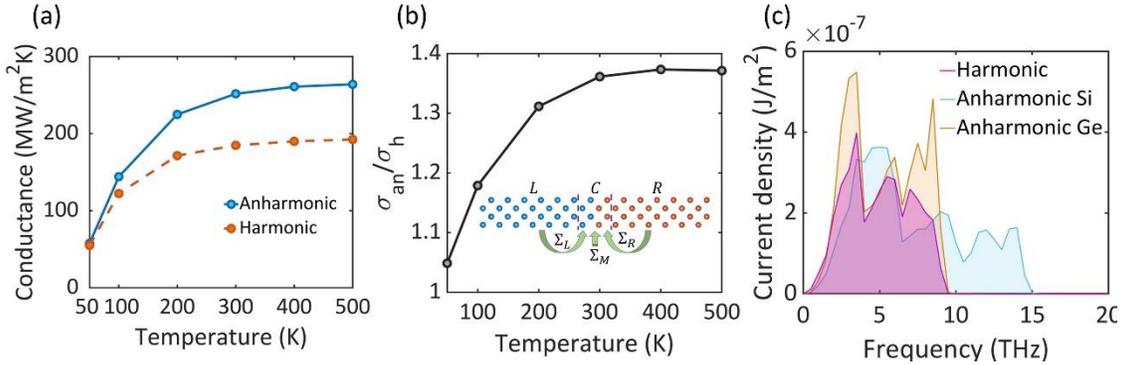

FIG. 1 (a) Conductance of Si/Ge interface, in the absence and presence of anharmonicity in the central region, as a function of temperature. (b) Conductance ratio versus temperature. Inset is the illustration of the system, where the semi-infinite left side (L) is Si meanwhile the right (R) is Ge. The central region (C) contains the interface. (c) Heat current frequency distribution of the harmonic case, in which heat flow is the same on both sides, and the anharmonic heat flow on Si and Ge side.

We represented the Si/Ge interface using the force constants of Si on both sides, the mass of Si on one side and the mass of Ge on the other side. The system is schematically depicted in the inset of Fig. 1(b). We obtained two-probe thermal conductance as a function of temperature in Fig. 1(a). Both harmonic and anharmonic conductance first rises with temperature, then stays constant because of fully excited modes following the Bose-Einstein distribution. The most striking feature here is that the anharmonic case gives significantly higher thermal conductance than the harmonic case. As shown in Fig. 1(b), anharmonicity can lead to nearly 40% enhancement in thermal conductance at room temperature. If one were to apply the four-probe conductance or include 4$^{th}$ order interactions, the ratio can be even higher. This is consistent with very recent observation from non-equilibrium molecular dynamics simulations[27].

To fully understand the enhancement of thermal interface conductance by anharmonicity, the frequency-resolved heat current density is shown in Fig. 1(c). In the harmonic case, the frequency distribution of heat current remains the same before and after the interface because only when there are phonon channels at the same frequency on both sides, phonons can propagate through the interface via elastic processes. After adding anharmonicity to the interface region, the current distribution on both sides is no longer the same and the total heat current is significantly increased. At the Si side, the phonon channels at higher frequencies than the cutoff frequency of Ge are opened and the allowed phonon channels cover the entire frequency range of pure Si. On the other hand, for the Ge side, even though the frequency range remains unchanged at the Ge cutoff frequency, the peaks become bigger because of the newly opened channel via anharmonicity. Here the anharmonic interface acts as a source of phonon-phonon coupling, which facilitates the energy communication among different phonon modes and assists the previously blocked phonons to propagate through the interface via inelastic scattering.

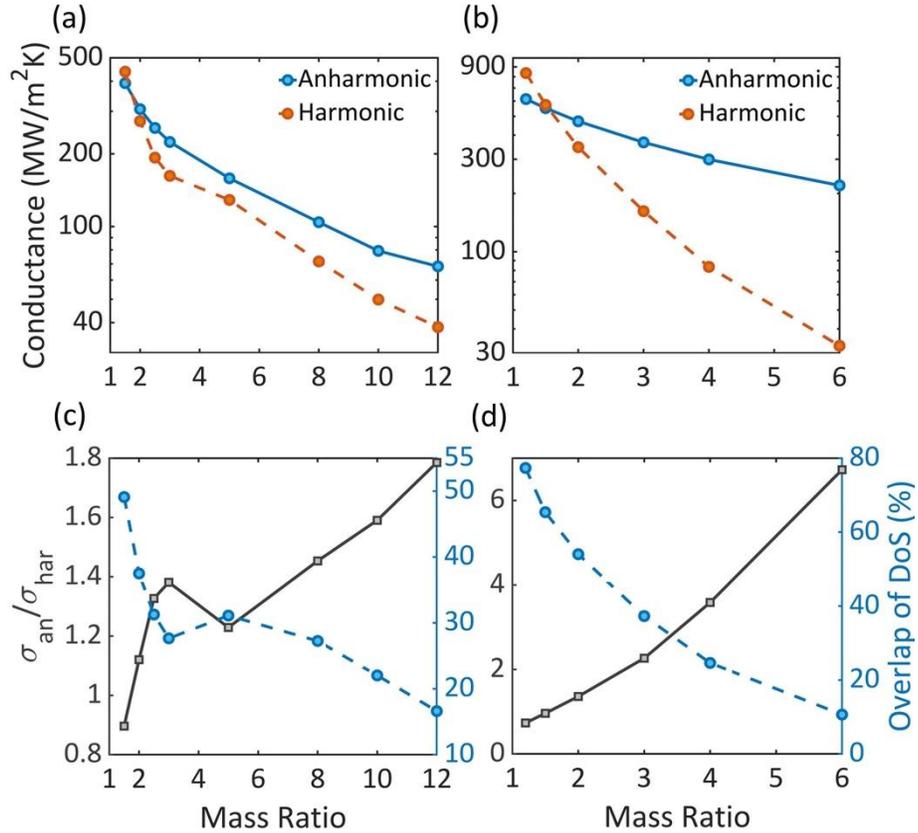

FIG. 2 (a)&(b) Conductance value as a function of mass ratio at 300K, for Si (a) and Al (b). The structure models used for anharmonic AGF are illustrated in the insets; (c)&(d) Thermal conductance ratio and overlap of versus mass ratio, for Si (c) and Al (d).

In order to explore the generality of the conductance enhancement for different interfaces, we performed further calculations on Si- and aluminum (Al)-based interfaces. We kept the left side to be Si or Al and varied the atomic mass on the other side. As mass ratio increase, the absolute values of thermal interface conductance reduces in both harmonic and anharmonic cases, as shown in Fig. 2(a) and 2(b). When the mass ratio is close to one, the interface behaves more like the pure material so including three-phonon scattering acts as a barrier for phonon transport to inhibit the otherwise perfect transmission. For Si-based interface, as illustrated in Fig. 2(c), the conductance ratio increases to a local maximum and then slightly decreases before the conductance ratio keeps increasing. In contrast, the conductance ratio of Al-based interface monotonically increases, as plotted in Fig.2(d). Apparently, the effect of anharmonicity is strongly system-dependent. Therefore, we need to understand the different conductance behaviors between Si and Al.

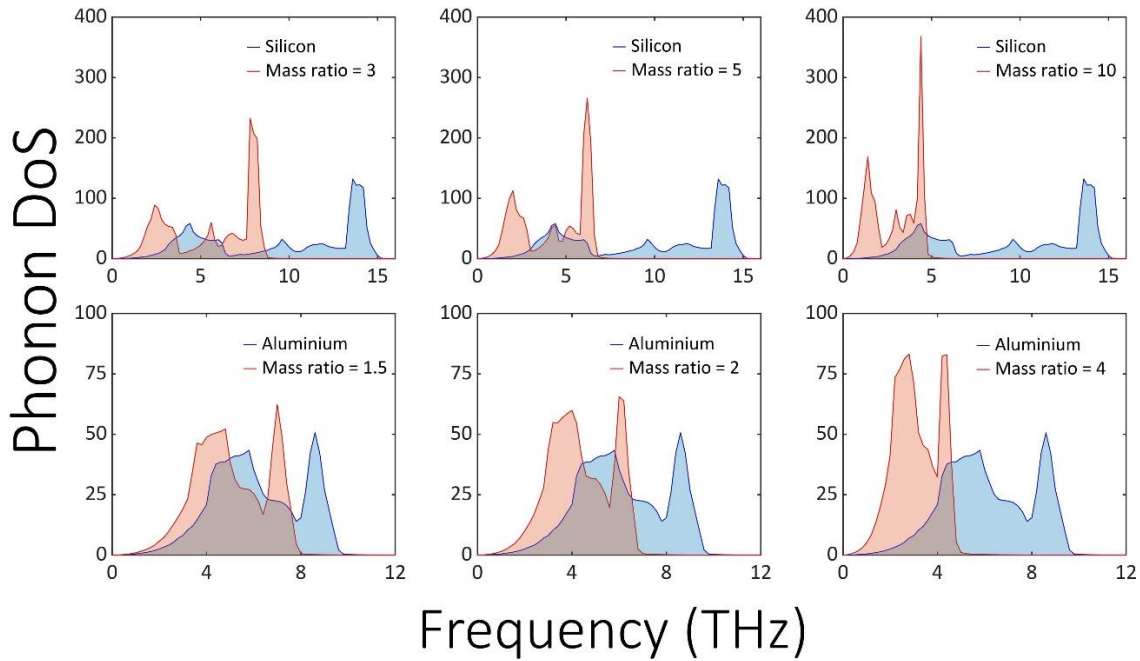

FIG. 3 DoS of the Si- or Al- based interface with different mass ratios; the blue one is Si or Al, while the red one is the DoS on the other side assigned with different mass.

The phonon density of states (DoS) on the two sides is shown in Fig. 3 to illustrate that the DoS mismatch will suppress allowed phonon channels in the harmonic case. The more disallowed channels blocked by mass difference in the harmonic regime, the more transmission channels anharmonicity can potentially enable. To illuminate the interplay of prohibited phonon channels and anharmonic effect, we used the overlap of DoS to quantify the available phonon channels in the harmonic case. As shown in Fig. 3, different materials have different DoS and this leads to different trends in overlap area as the mass ratio increases. Nevertheless, the thermal conductance ratio vs. mass ratio changes in exactly the opposite trend as the overlap of DoS for both Si and Al cases, as shown in Fig. 2 (c) & (d). In other words, if there are more channels for anharmonicity to open up, the ratio of anharmonic vs. harmonic conductance is larger.

*Conclusion*—In summary, we established a rigorous anharmonic AGF formalism to include the anharmonicity at interfaces for the 3-D structure with first-principles force constants. To do so, we developed a new Fourier decomposition method for the 3$^{rd}$ order anharmonic tensors and introduced the three-phonon scattering at solid-solid interfaces via the many-body self-energy in Fourier space. We revealed the significantly increased thermal conductance at Si/Ge interface. We attributed the thermal transport enhancement to the phonon channels opened up by inelastic phonon scattering. We further quantitively showed that the smaller DoS overlap at both the Si-based and Al-based interfaces, the more significant role anharmonicity plays. By overcoming the long-standing challenges of including anharmonicity into AGF for 3-D structures, this work brings AGF method to a new height. The application scope of AGF has been remarkably extended, and the importance of anharmonicity at the interface has been highlighted in this Letter.

Acknowledgement

This work is sponsored by the Department of the Navy, Office of Naval Research under ONR award number N00014-18-1-2724. This work used the Extreme Science and Engineering Discovery Environment (XSEDE), which is supported by National Science Foundation grant number ACI-1053575.